\documentclass[traditabstract]{aa} 
\usepackage{graphicx}
\usepackage{txfonts}
\usepackage{soul} 
\begin{document}
   \title{The Inner Halo of M87:  A First Direct View of the Red-Giant Population}


   \author{Sarah Bird
          \inst{1}
          \and
          William E. Harris
          \inst{2}
	  \and
	  John P. Blakeslee
          \inst{3}
	  \and
          Chris Flynn\inst{1,4,5}
          }

\authorrunning{Bird et al.}
        
   \institute{Tuorla Observatory, University of Turku, V\"{a}si\"{a}l\"{a}ntie 20,
FI-21500 Piikki\"{o}, Finland,
              \email{sarbir@utu.fi, cflynn@utu.fi}
         \and
             Department of Physics and Astronomy, McMaster University, Hamilton L8S 4M1, Canada,
             \email{harris@physics.mcmaster.ca}
	\and
	     Herzberg Institute of Astrophysics, National Research Council of Canada,
             Victoria BC V9E 2E7, Canada,
             \email{john.blakeslee@nrc.ca}
        \and
Finnish Centre for Astronomy with ESO (FINCA), University of Turku,
V\"{a}si\"{a}l\"{a}ntie 20, FI-21500 Piikki\"{o}, Finland
        \and
           Department of Physics and Astronomy, University of Sydney, Sydney,
Australia
     }

   \date{Received (date)/ Accepted (date)}

 
  \abstract{An unusually deep $(V,I)$ imaging dataset for the Virgo supergiant
    M87 with the Hubble Space Telescope ACS successfully resolves its brightest
    red-giant stars, reaching $M_I(lim) = -2.5$.  After assessing the
    photometric completeness and biasses, we use this material to estimate the
    metallicity distribution for the inner halo of M87, finding that the
    distribution is very broad and likely to peak near [m/H] $\simeq -0.4$ and
    perhaps higher.  The shape of the MDF strongly resembles that of the inner
    halo for the nearby giant E galaxy NGC 5128.  As a byproduct of our study,
    we also obtain a preliminary measurement of the distance to M87 with the
    TRGB (red-giant branch tip) method; the result is $(m-M)_0 = 31.12 \pm
    0.14$ ($d = 16.7 \pm 0.9$ Mpc). Averaging this
    result with three other recent techniques give a weighted mean 
    $d(M87) = (16.4 \pm 0.5)$ Mpc. }


   \maketitle
%

\section{Introduction}

A key route to understanding the early enrichment history of large galaxies is
through direct, star-by-star measurement of the metallicity distribution
function (MDF) of its halo stellar population.  Direct resolution of the
old-halo stars with ground-based imaging is, however, generally restricted to
the relatively small number of galaxies in the Local Group and not much beyond.
The Hubble Space Telescope cameras with their superior spatial resolution and
depth are able to extend this type of study considerably further, bringing many
more galaxies of all types within reach.

The rarest and biggest types of galaxies are the giant elliptical (gE) types,
and to date, very few direct measurements of their MDFs have been made through
their halo-star populations.  Data of this type have been obtained with HST
imaging only for the nearest gE galaxies, namely NGC 5128 at $d = 3.8 \pm 0.1$
Mpc (Harris et al. \cite{har99}, \cite{har00}, \cite{har02}, \cite{har10},
Rejkuba et al.~\cite{rej04}) and NGC 3379 at $d = 10.5 \pm 0.8$ Mpc (Gregg et
al. \cite{gre04}, Harris et al. \cite{har07}). To these we may add the MDF
study of the intermediate-luminosity Leo elliptical NGC 3377, at $d=10.8 \pm
0.5$ Mpc (Harris et al.~\cite{har07a}).  Mouhcine et al. (\cite{mou05b},
\cite{mou07}) have used HST images to directly measure the halo-star MDF of
nine spiral galaxies, finding a correlation between overall luminosity and the
mean halo metallicity, and suggest that halos of massive spirals may not be
dominated by metal weak stars.  For this kind of work, the Virgo cluster of
galaxies at $d \simeq 16$ Mpc is an especially tantalizing arena because it
holds a rich collection of all types of galaxies including many large
ellipticals.  Among all of the Virgo members, however, M87 (NGC 4486) can be
claimed to be the most intriguing target because it is the nearest example we
have of a centrally dominant cD-type galaxy, which is likely according to the
current galaxy formation framework (e.g. Thomas et al. \cite{tho05}; Renzini
\cite{ren06} for comprehensive discussions) to have undergone its major
star-forming period very quickly in the early universe.

In this paper, we take advantage of an unusual dataset in the HST Archive to
probe the inner-halo stellar population of M87 -- that is, its fundamental
population of old red-giant stars -- for the first time.  As will be seen in
the discussion below, our ability to gauge the MDF accurately is affected by
the high level of photometric crowding in the images.  Despite its limitations,
however, it is good enough to permit some preliminary steps towards measuring
the stellar MDF.  In the following sections, we describe the raw data and our
preprocessing; the photometric methods; measurement of the red-giant-branch tip
and distance calibration; and finally a preliminary discussion of the
metallicity distribution function.

\section{The Dataset and Preliminary Processing}

The raw images we use in this study are from HST program GO-10543 (PI Baltz)
and were taken with the ACS (Advanced Camera for Surveys) Wide Field Channel in
the $F606W$ (wide $V$) and $F814W$ (wide $I$) filters.  The $3.4' \times 3.4'$
field, corresponding to $16 \times 16$ kpc linear scale at the Virgo distance,
was centered on M87.  The exposures were taken in 61 separate spacecraft
visits, extending over a 71-day period from 2005 Dec 24 to 2006 March 5.  These
many exposures were originally designed to search for microlensing events. In
$F814W$ there are 205 images totalling 73800 seconds, while in $F606W$ there
are 49 images totalling 24500 seconds.  Co-adding the individual exposures in
each filter yields an unusually deep pair of images at higher scale and 
a certain amount of gain
in spatial resolution.  Composite exposures of these images have already been used for
three separate studies of the population of $\simeq 2000$ globular clusters
that lie within the area of this ACS pointing (Waters et al.~\cite{wat09}, Peng
et al.~\cite{peng09}, Madrid et al.~\cite{mad09}).

If the distance to M87 is 16 Mpc (see below), the expected magnitude of the
TRGB (red-giant branch tip) at $M_I = -4.05$ (Rizzi et al.\ \cite{riz07}) is
$I_{TRGB} \simeq 27.0$.  As will be seen below, the limiting $V,I$ magnitudes
of these co-added exposures are easily capable of directly resolving the TRGB
and delineating the color-magnitude distribution (CMD) of the brightest RGB
stars.  The CMD can, in turn, be used to derive the MDF, as well as a direct
calibration of the distance to M87 through the luminosity of the RGB tip.

For the purposes of this study, we carried out a new combination of all the raw
exposures in each filter through the APSIS software (Blakeslee et
al.~\cite{bla03}), which performs accurate image registration, cosmic-ray
rejection, and distortion correction with drizzle.  Many hundreds of sources
were used for measuring the offsets and small rotations of the individual
exposures, yielding an accuracy of the registration of the raw images better
than 0.01~px.  The unusually large numbers of exposures in each filter
permitted subpixel resampling; we chose to construct final combined science
images with a scale 0\farcs025~px$^{-1}$ (half the native pixel size of the
camera), and used the drizzle ``point'' (delta-function) interpolation kernel,
which is equivalent to a square kernel with a pixfrac parameter of zero.  Note
that in the previous globular cluster analyses of Peng et al.\ (\cite{peng09})
and Madrid et al.\ (\cite{mad09}) we generated combined images with the
Gaussian kernel and a scale of 0\farcs035~px$^{-1}$.  However, for the present
study, we decided to rebuild the final science frames to gain the best possible
effective resolution for faint-star photometry. 

\begin{figure*}
\centering
\includegraphics[height=7.0cm]{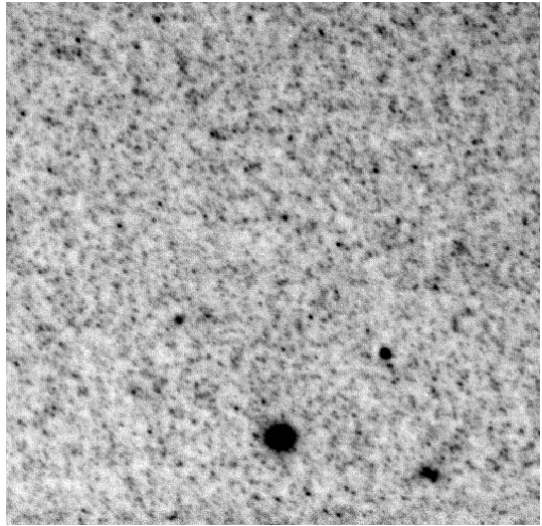}
\caption{A $12.5'' \times 12.5''$ portion of the composite $F814W$ image of the
  M87 halo, demonstrating its resolution into stars.  Three globular clusters
  are seen in the lower half of the picture.  This section is located in the
  extreme southeast corner of the ACS field, $\simeq 150''$ or 12.1 kpc from
  the center of M87.  Stellar crowding increases rapidly toward the galaxy
  center but is manageable at this location.}
\label{snapshot}
\end{figure*}

\begin{figure*}
\centering
\includegraphics[height=7.0cm]{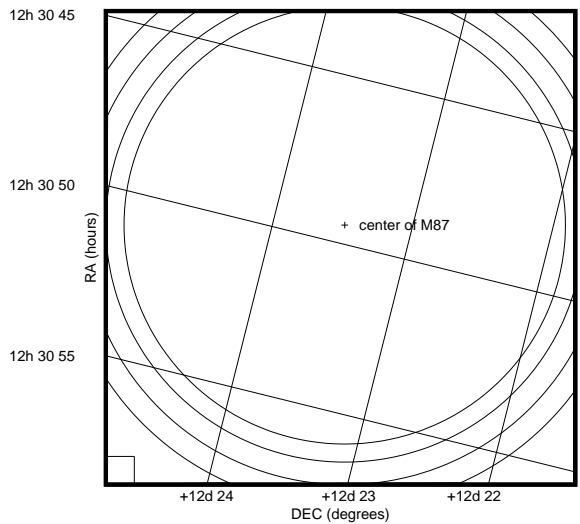}
\caption{Full ACS field of view. Lines corresponding to declination versus
    right ascension are drawn. Four annuli mark regions used in the photometry
    at $97'' - 105'', 105'' - 115'', 115'' - 125'',$ and $125'' - 155''$, and
    are drawn around the center of M87. Note that the 155'' circle does not
    show, because it is coincident with the bottom left corner of the ACS
    image. The central circle and the first two annuli are not used for final
    analysis due to crowding. Figure \ref{snapshot} is the portion marked by a
    square in the lower left corner.}
\label{cartoon}
\end{figure*}

A portion of our final science image in $F814W$, illustrating the degree of
crowding in an outer corner of the field, is shown in Figure \ref{snapshot}.
The excised sample shown in the Figure is located in the southeast corner of
the ACS field, at $12~30'~58.576''$ and $+12^{\circ} 24' 39.00''$, at a
projected galactocentric distance of about $150''$ or 12.1 kpc.  The mean
surface brightness of the region in Figure \ref{snapshot} in V and I bands is
$\mu_V=22.4$ mag/sq arcsec and $\mu_I=21.2$ mag/sq arcsec, from the
  surface photometry of Kormendy et al.~\cite{kor09} (see also Zeilinger et
al.~\cite{zei93}).  The location of this field is shown in Figure
\ref{cartoon}.

Another similarly deep probe of the RGB population in the Virgo region has been
published by Williams et al.~(\cite{wil07}).  Their HST/ACS target field (from
program GO-10131) is located in the intracluster region between M87 and M86,
approximately 180 kpc projected distance from each of these two giants. Their
data consist of 26880 s exposure time in $F814W$ and 63440 s in $F606W$, very
similar total integration but with the opposite color emphasis than our central
M87 field has.  As Williams et al. discuss, the stellar population in this
intracluster field is predominantly metal\emph{-poor}, and is likely to have a
large mixture of material tidally stripped from many Virgo members.  By
contrast, the M87-centered field we analyze here should provide a relatively
pure sample of the inner-halo RGB stars in this supergiant E alone and can be expected to be more metal-rich on average.

\section{Photometry}

To define the point-spread function (PSF) empirically for stellar photometry,
we selected 22 moderately bright, isolated stars located across the ACS field
and used these same stars for both filters (only about 40 bright stars in total
are clearly distinguishable in the entire field among the $\sim2000$ globular
clusters; see Madrid et al. 2009).  Our target RGB halo stars are far fainter
than these PSF stars, and the internal measurement uncertainties are dominated
completely by sky noise and crowding, so the fine details of the PSF shape are
not critical for this purpose.  Nevertheless, we found no detectable
differences in PSF size as a function of location in the field after the APSIS
distortion corrections (see also Madrid et al.~for further discussion).  The
FWHM of the PSF we obtained is 3.2 px ($0.080''$) in $F606W$ and 3.4 px
($0.085''$) in $F814W$, yielding a small but useful gain in effective
resolution compared with the nominal $0.1''$ resolution of the individual
exposures.

Across almost all the area of this M87-centered ACS field the level of stellar
crowding is, to say the least, severe.  It is only in the outermost corners
that crowding is low enough to permit star-by-star identification and
photometry by normal techniques.  Once the PSF was established, we masked off
all the area within 3900 px (7.9 kpc or $97''$) projected galactocentric
distance, essentially leaving only the four corners.  This outer measured
region is marked in Fig.~\ref{cartoon}.  To carry out the photometry, we used
the IRAF \emph{daophot} codes (see Stetson \cite{ste87}) with a 2-pass sequence
of \emph{find/phot/allstar}.  For \emph{allstar} the PSF fitting radius was
adopted as 1.5 px for the final runs to reduce crowding effects as much as
possible, though we found that slightly larger values up to 2.5 px did not give
significantly different results.  Correction of the PSF-fitted instrumental
magnitudes to large-aperture radius was done by curve-of-growth measurements
from the same set of bright stars that were used to define the PSF.  Lastly,
these aperture-corrected instrumental magnitudes were converted to $F606W,
F814W$ magnitudes and then into $(V,I)$ with the current ACS/WFC filter
zeropoints and conversion relations in the VEGAmag system (see Sirianni et al.
\cite{sir05} and the ACS web
pages\footnote{http://www.stsci.edu/hst/acs/analysis/zeropoints}),

\begin{eqnarray}
F606W = - {\rm 2.5 log}(f_V) + 26.421 \\ 
F814W = - {\rm 2.5 log}(f_I) + 25.536 \\
V = F606W + 0.236 (V-I) \\ 
I = F814W - 0.002 (V-I) 
\end{eqnarray}

\noindent where $f_V, f_I$ are the flux rates in counts per second for large-aperture
photometry.  The dominant
uncertainty in the calibration is in the correction from the small radii (2 px)
that we are forced to use because of the high degree of crowding, to
``infinite'' large-aperture radius through the curves of growth.  From these we
believe the true external uncertainty in the photometric zeropoints to be
$\pm0.1$ mag in each filter.

\begin{figure*}
\centering
\includegraphics[height=11.0cm]{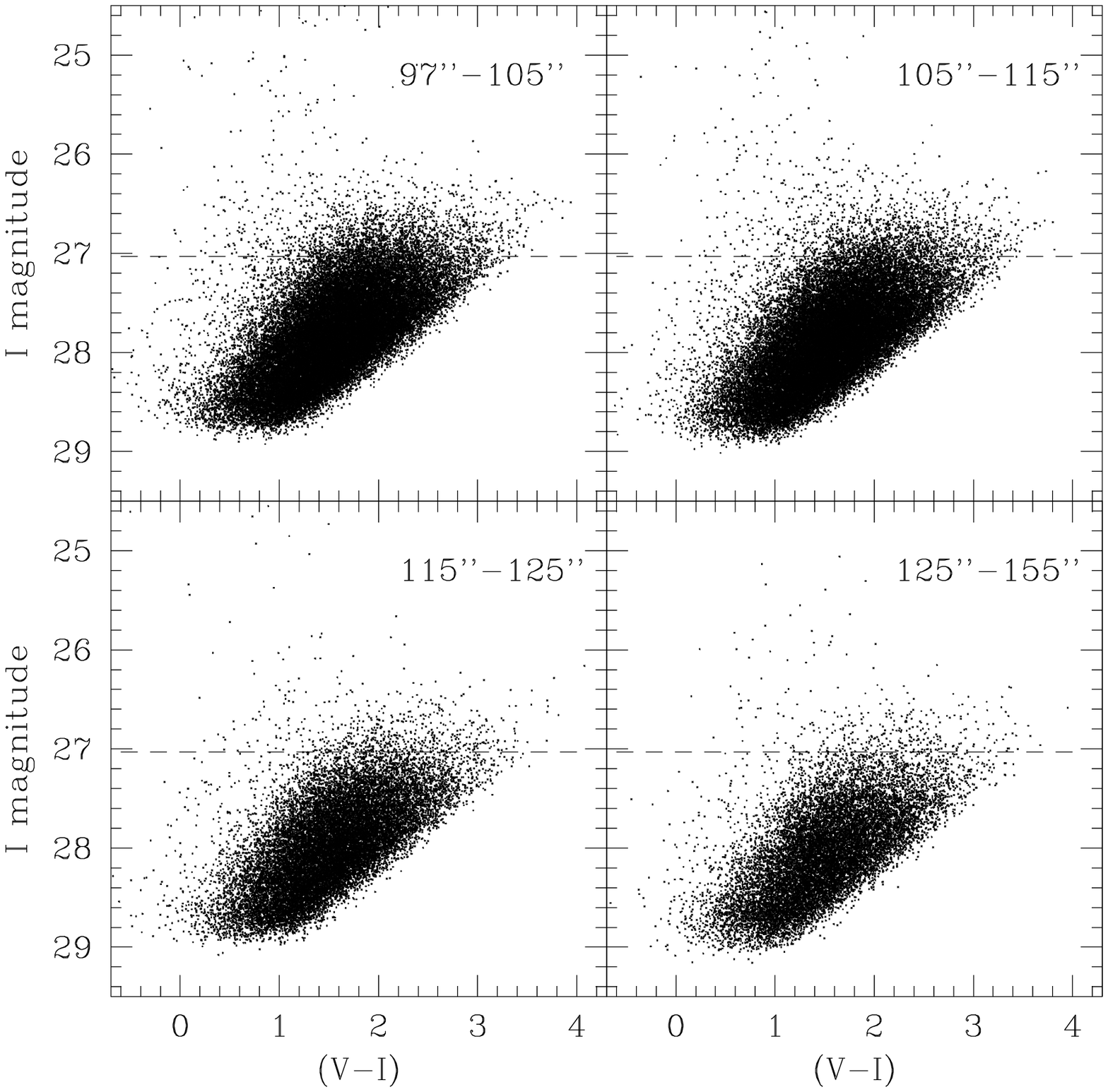}
\caption{The color-magnitude distribution ($I$ vs.~$(V-I)$) for the measured
  stars in the M87 inner halo.  Data are plotted in four concentric annuli in
  projected galactocentric distance.  These annuli cover a total radial range
  in $R_{gc}$ from 7.9 kpc to 12.5 kpc.  The
  horizontal dashed line in each panel shows the \emph{predicted} level of the
  RGB tip at $M_I = -4.05$ (see text).  }
\label{cmd4}
\end{figure*}

Over the outer region defined above, our final data consist of 105466 measured
stars.  The color-magnitude array for these is shown in Figure \ref{cmd4},
subdivided into the four annular zones that are shown in Fig.~\ref{cartoon}. To
facilitate comparison between zones, the expected level of the RGB tip
luminosity for \emph{metal-poor} ([Fe/H] $\lesssim -0.7$) RGB stars at $M_I =
-4.05$ (Rizzi et al.~\cite{riz07}) is shown in each panel as the dashed line.
For our present purposes, to minimize crowding effects as much as possible we
restrict our following discussion to the two outermost zones ($R_{gc} =
115''-155''$), containing 33890 measured stars.

\begin{figure*}
\centering
\includegraphics[height=8.0cm]{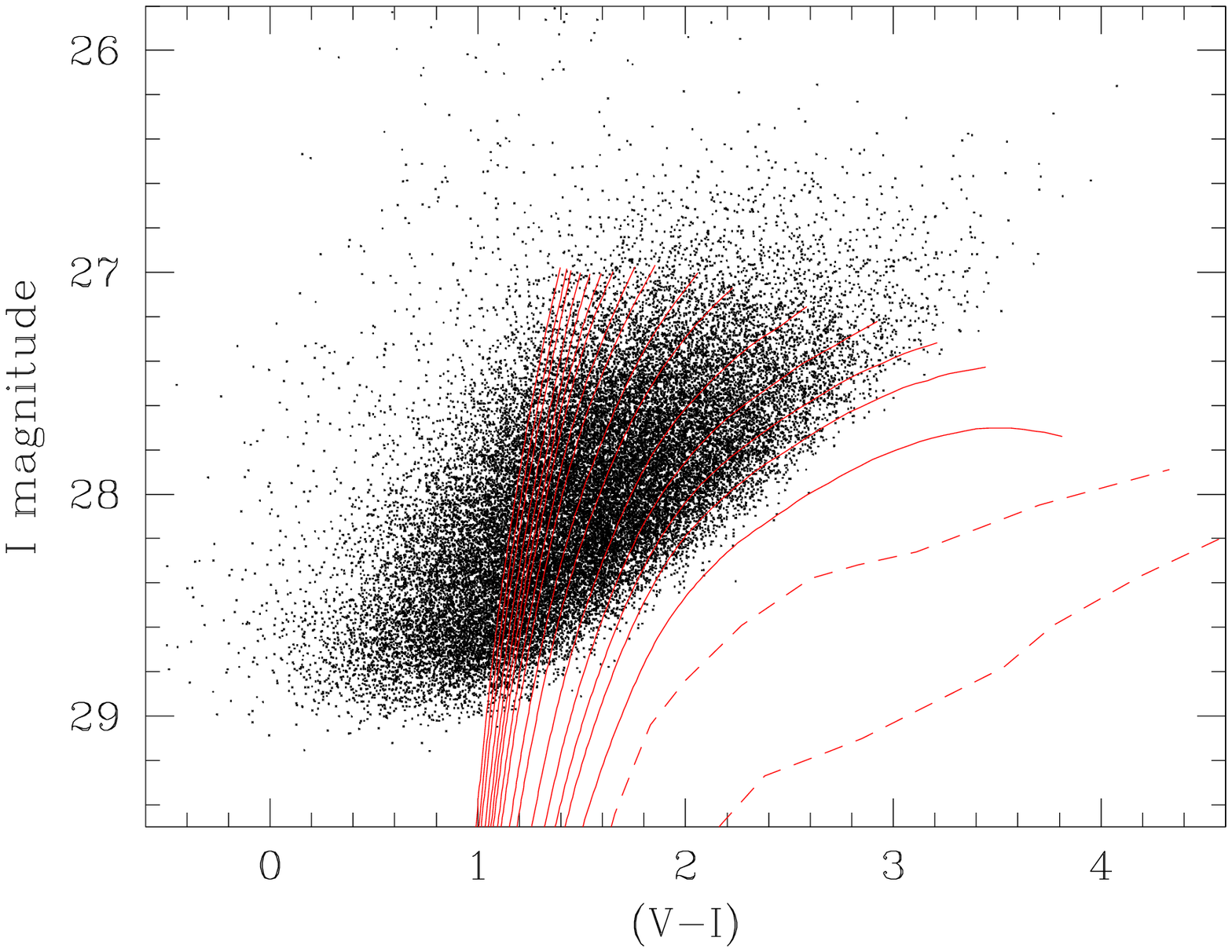}
\caption{Color-magnitude distribution ($I$ vs.~$(V-I)$) for the measured stars
  in the outer radial range ($115'' - 155''$) of our ACS field.  The
  superimposed solid lines are fiducial 12-Gyr RGB tracks from the VandenBerg
  et al.  \cite{vdb00} model library, in steps of roughly 0.1 dex from log
  $(Z/Z_{\odot}) = -2.0$ to $-0.1$.  These are supplemented with empirical
  tracks for two high-metallicity values at log $(Z/Z_{\odot})$ = +0.16 and
  +0.4 (dashed lines; see Harris et al. 2002).  }
\label{cmd_fiducials}
\end{figure*}

The RGB spans a broad color range, suggesting a large range in metallicity, or
very large internal photometric scatter, or both.  Our ability to see the true
metallicity spread of the RGB tip is limited by the $F606W$ exposures, which
set the very distinct red-edge cutoff to the data at $V \lesssim 30$ (for
similar cases, see the NGC 3379 or NGC 5128 studies of Harris et
al. \cite{har02}, \cite{har07} where the most metal-rich part of the population
is cut off).  A more specific demonstration of this point is shown in Figure
\ref{cmd_fiducials}, where the combined photometry for the two outermost annuli
is plotted along with fiducial tracks for 12-Gyr-old RGB stars over the
metallicity range [Fe/H] $= -2.3$ to $+0.4$.  These tracks are the same ones
used in previous studies of NGC 5128 and NGC 3379 (Harris et
al. \cite{har02,har07}) and are drawn primarily from the model library of
VandenBerg et al. (\cite{vdb00}).  At the right-hand edge of the CMD, the
detection limits set by the $V$ filter prevent us from measuring any stars more
metal-rich than [m/H] $\simeq -0.2$ (i.e. stars in the range of the three
reddest tracks in the model grid).  In addition, at levels $I \gtrsim 27.6$ the
increasing photometric measurement uncertainties produce an \emph{asymmetric}
scatter in the observed colors: objects measured too blue by random errors fall
well to the blue side of the most metal-poor model track; contrarily, any that
are measured too red by similar amounts fall below the $V$ limits and do not
appear on the CMD.  These arguments suggest that the MDF derivation needs to be
restricted to only the very brightest section of the RGB to minimize biasses
and excessive random spread. We will quantify this approach below after a brief
discussion of the TRGB and distance calibration.

\section{Determining the TRGB and Fiducial Distance}

Although the main focus of our M87 study is the metallicity distribution, the
data are also suitable for a preliminary estimate of the tip of the red-giant
branch (TRGB).  The bright-end termination of the red giant branch has proven
to be an extremely effective standard candle for distance measurement to nearby
galaxies (well reviewed by Rizzi et al. \cite{riz07}).  Observationally it has
the strong advantages that only single-epoch imaging is needed; that it is used
for halo stars where internal differential reddening is usually not an issue;
and that it rests soundly on a well defined and well understood breakpoint in
the evolution of old, low-mass stars that are universally present in most
galaxies.  In addition, it is most easily employed in the $I$ band where the
level of the TRGB is nearly flat for the generous metallicity range [Fe/H]
$\lesssim -0.7$ (see Rizzi et al. and papers cited there).

\begin{figure*}
\centering
\includegraphics[height=7.0cm]{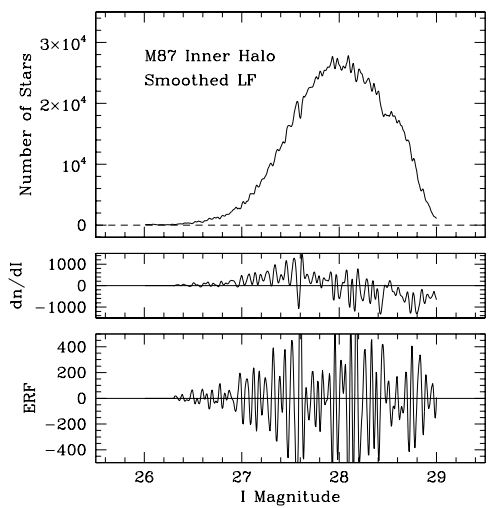}
\caption{Definition of the luminosity function for the M87 stars and the TRGB.
  Top panel shows all stars in the outer radial range ($115'' - 155''$) of our
  ACS field, smoothed with a 0.01-mag Gaussian kernel.  Middle and bottom
  panels show the numerically derived first and second derivatives of the LF.
}
\label{trgb}
\end{figure*}

The TRGB, which physically represents the helium-ignition point along the
evolutionary tracks of the red-giant stars, is empirically defined by a sudden
rise in the luminosity function (LF) of the halo stars.  For recent examples of
the sharp, near-ideal TRGBs that result from uncrowded photometry and
statistically large samples of stars, see McConnachie et al.~(\cite{mcc05}),
Mouhcine et al.~(\cite{mou05a}), or Harris et al.~(\cite{har07}).  However, our
M87 photometry is unavoidably in a different regime: because of the crowding
and substantial star-by-star photometric uncertainties (which are already as
large as $\pm0.36$ mag at the top of the RGB, as detailed in the next
section), the TRGB does not appear here as a sharp onset in the LF but rather a
smoothed-out rise.  In Figure \ref{trgb} we show the LF for the measured stars
in the outer radial range ($115'' - 155''$) of our M87 ACS field.  Here, the
data are smoothed with a Gaussian kernel of $\sigma = 0.01$ mag, though the
result is insensitive to the particular smoothing width.  The numerical first
derivative $(dn/dI)$ and second derivative $(d^2n/dI^2)$ or ``edge response
filter'' ERF are shown below the smoothed LF.

A prominent increase in the ERF amplitude first appears at $I = 26.98 \pm 0.03$
(internal uncertainty), which we identify as the TRGB.  This apparent magnitude
must be adjusted for a systematic photometric error (see next section) of
$\Delta I = 0.12$ mag, yielding a true TRGB level $I(TRGB) = 27.10$.  Adopting
$M_I(TRGB) = -4.05 \pm 0.10$ (Rizzi et al. \cite{riz07}), we obtain an apparent
distance modulus $(m-M)_I = 31.15 \pm 0.14$, where we have now included
the $\pm 0.1-$mag uncertainty in the photometric zeropoint (see above).
Subtracting $A_I = 0.03$ as listed in the NED then gives $(m-M)_0 = 31.12 \pm
0.14$ and a true distance to M87 of $d = (16.7 \pm 0.9)$ Mpc.

\begin{figure*}
\centering
\includegraphics[height=7.0cm]{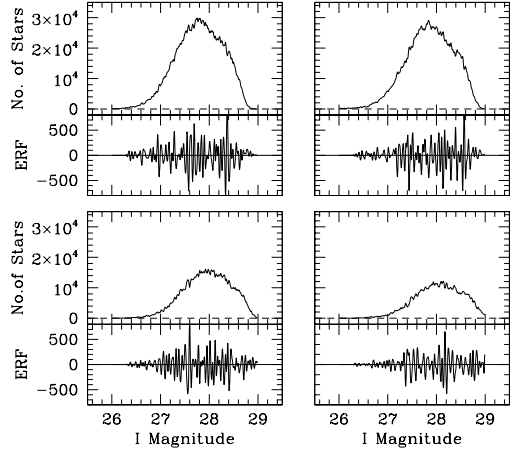}
\caption{The luminosity function and edge response filter calculations for the
  four radial zones defined in Fig.~\ref{cmd4} above.  All have been determined
  in the same way as in the previous figure.  }
\label{edge4}
\end{figure*}

As a secondary test of the degree to which crowding alone affects our result,
we show in Figure \ref{edge4} the smoothed LF and ERF for the four separate
radial zones defined in Figs.~\ref{cartoon} and \ref{cmd4}.  Aside from
differences in the total numbers of stars, no significant offsets are seen in
the estimated TRGB from field to field.

Given that the ERF technique is difficult to apply in such high crowding
regimes, we estimated the TRGB level with an independent approach, similar to
the ones used by Harris et al. (\cite{har98}) and McConnachie et al.
(\cite{mcc04}).  Here, the LF of the observed stars is modelled as the sum of
an AGB and RGB for an adopted distance modulus, convolved with the photometric
errors, and compared to the observed LF. The RGB is assumed to have a step
function onset at the TRGB, rising exponentially to fainter magnitudes; the AGB
is assumed to start one magnitude brighter than the TRGB, rising exponentially
but with lower amplitude than the RGB. (In practice, the AGB can be thought of
as the sum of the true old-AGB stars plus blended pairs from the RGB.) The
exponential slopes are set to $\Delta\log\,N/\Delta I = 0.96$, and the
normalisation of the AGB to the RGB is a free parameter. We convolve this model
by the photometric error and completeness as functions of $I$ magnitude
(determined from our tests with artificial stars), and compare it to the
observed LF. We found good fits to the observed LF for an M87 TRGB at
$I = 27.1 \pm 0.2$ and an AGB-to-RGB ratio at the TRGB of $\approx 5 - 10$ percent.
The AGB/RGB ratio plays only a minor role in the fitting, which is most
sensitive to I(TRGB).  The fitting corroborates $I(TRGB) = 27.1$ derived
from the ERF method above.

The stars lying above the RGB tip will be a combination of (a) field
contamination from both foreground stars and faint, small background galaxies;
(b) stars that are genuinely part of the M87 halo/bulge population, either
young evolved stars or old AGB stars in highly evolved states including
long-period variables; and (c) accidental blends of two or more normal RGB
stars.  The field contamination component $N_f$ is independent of the number
$N_{RGB}$ of M87 stars; the number of AGB stars goes in proportion to
$N_{RGB}$; and the number of blends varies as $N_{RGB}^2$.  For our two outer
zones, there are 761 stars in the $0.7-$mag region just above the TRGB, versus
4467 stars in the $0.5-$mag region just below it.  For field contamination,
comparison with the photometry of Williams et al.~(\cite{wil07}) and Madrid et
al.~(\cite{mad09}) suggests that $\la 40$ foreground stars are present, a
negligible contribution.  The number of faint, small and misidentified
background galaxies is harder to assess without an adjacent control field to
rely on, but comparison with the similar ACS $(V,I)$ Leo field data of Harris
et al.~(\cite{har07a}) suggests very conservatively $N_{gal} \la 50$, also
negligible.

For the NGC 3379 halo field, Harris et al.~(\cite{har07}) found that the number
of LPVs was 2.5\% of the number of stars in the brightest 0.5 mag of the RGB.
Scaling from that fraction then indicates we should expect $N_{LPV} \simeq 110$
in our two outer zones.  These plus the field contaminants then account for
about a quarter of the CMD stars above the TRGB.  There is no evidence for
any young population of stars, which if present would produce an upward
extension of the blue side of the RGB (see Williams et al.~\cite{wil07} for
an example).  

Most or all of the remainder is likely to be due to blends: we expect (see
Harris et al.~\cite{har07a})
\begin{equation}
N_{blend} \, \simeq \, {N_{\star}^2 \over 2} {{a} \over A}
\end{equation}
where $N_{\star}$ is the number of stars in the field capable of generating a
blended pair brighter than the TRGB, $a$ is the area of one resolution element,
and $A$ is the area of the field.  We take $A = 1.42$ arcmin$^2$ for the area
of the outer two annuli enclosed within the field; $a = 7.8 \times 10^{-3}$
arcsec$^2$; and $N_{\star} \simeq 16000$ for the number of stars between $I=27$
and 28, the top magnitude of the RGB.  These parameters predict about 200
blends lying above the TRGB.  In fact this estimate is only a lower limit,
because (as discussed in the next section) crowding-driven incompleteness of
the photometry affects even the range $I < 28$ and $N_{\star}$ could be quite a
bit larger than our directly measured count.  Though admittedly provisional,
our conclusion is that the great majority of the objects brighter than the TRGB
can be accounted for by the combination of the three normal factors given
above.

One useful comparison study of the distance to M87 is the photometry of the
intracluster stars in Virgo by Williams et al. (\cite{wil07}).  Although their
target field is far from M87, it should in principle have a reasonably similar
TRGB level given that M87 is near the physical center of the Virgo environment.
Although they do not specifically derive the TRGB, their CMD for the
intracluster stars clearly shows $I(TRGB) \simeq 26.9 - 27.0$, close to our
value for M87 itself.

Other indirect, but still quite relevant, TRGB-based calibrations of the Virgo
distance include studies of a small number of its dwarf ellipticals.  Harris et
al.~(\cite{har98}) obtained $(m-m)_0 = 30.98 \pm 0.20$ for VCC1104 = IC3388,
while Caldwell (\cite{cal06}) obtained $(m-M)_0 = 31.03 \pm 0.1$ for seven
other dwarfs.  All of these are consistent with membership in the Virgo
cluster, and with our result for M87.

The only other \emph{direct} distance calibrations for M87 that rely on
resolved stars are (a) the observations of two novae (Pritchet \& van den Bergh
\cite{pri85}), from which they obtain only the extremely uncertain range of
30.4 -- 32.4 for the distance modulus; and (b) the planetary nebula luminosity
function (PNLF) by Ciardullo et al. (\cite{cia98}).  After adjustment of their
adopted zeropoint $M^{\star}(\lambda 5007) = -4.54$ upward to $M^{\star} =
-4.67$ to reflect a change in the fiducial M31 distance (see Harris et
al.\ \cite{har10}), the PNLF modulus is $(m-M)_0 = 30.92 \pm 0.16$.  The
modulus is 0.2 mag smaller than our TRGB value, though the two methods agree
within their combined uncertainties.

The largest and most precise compilation of \emph{relative} distances for the
Virgo galaxies is by Blakeslee et al.~(\cite{bla09}; see also Mei et
al. \cite{mei07}), from the surface brightness fluctuations (SBF) technique.
SBF represents a ``secondary'' standard-candle method because it relies for its
calibration on nearby galaxies whose distances are known beforehand from
resolved-star methods, although theoretical calibration via stellar population
models is also possible.  Blakeslee et al.\ (\cite{bla09}) determine $(m-M)_0 =
31.11 \pm 0.08$ for M87 in particular, starting from a \emph{mean} distance
modulus of $31.09 \pm 0.03$ for the Virgo cluster as a whole.  This in turn is
based on the Cepheid calibration of the SBF method given by Tonry et
al.\ (\cite{tonry2001}), corrected following Blakeslee et al.\ (\cite{bla02})
to the final set of Key Project Cepheid distances (Freedman et
al. \cite{freedman2001}).

Our TRGB distance measurement, and the earlier PNLF measurement, are gauges of
the M87 distance that rely much less on Cepheids.  The PNLF zeropoint, for
example, relies more heavily on the fiducial distance to M31, for which close
agreement now holds among several standard candles including RR Lyrae stars,
TRGB, and Cepheids.  Insofar as all these methods start from a common,
fundamental Local Group distance scale that has been carefully assembled by a
combination of many standard candles, the $\sim \pm0.2-$mag variances noted
above are reflective of the small discrepancies that can emerge from the
different methods in ways that are often still hard to pinpoint and that are
not even consistent from one galaxy to another.  For a more extensive
discussion on another system (NGC 5128) where several stellar candles can be
accurately compared, see Harris et al. (\cite{har10}).

Another direct, though somewhat less precise, comparison method of interest is
the linear size distribution of globular clusters (GCs), which has been
developed by Jord\'an et al.\ (\cite{jor05}) into a standard-ruler technique.
The key quantity is the peak of the GC half-light radius distribution,
normalized to host galaxy size and calibrated via the Milky Way GCs.  Using the
Jord\'an et al. data for M87 and their calibration, we obtain $d = (16.4 \pm
2.3)$ Mpc or $(m-M)_0 = 31.07 \pm 0.30$.

Combining the four methods listed above (TRGB, PNLF, GC sizes, SBF/Cepheids),
we obtain a weighted average distance modulus $\langle m-M \rangle_0 = 31.08
\pm 0.06$ for M87, or $d = (16.4 \pm 0.5)$ Mpc.  A more precise TRGB distance
especially could be obtained very straightforwardly with halo-star photometry
in a less crowded region, and would in our view be the most effective way to
calibrate the distance to this important galaxy.

\section{Tests of the Photometry}

\begin{figure*}
\centering
\includegraphics[height=8.0cm]{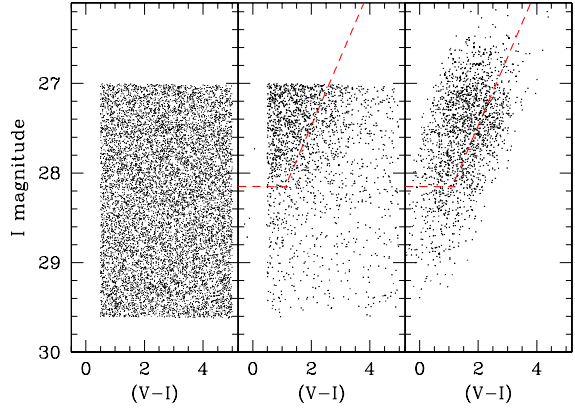}
\caption{Results of the first artificial-star set of tests.  \emph{Left panel}
  shows the color-magnitude distribution of the artificial stars inserted into
  the region of the image described in the text.  \emph{Middle panel} shows the
  stars from the left panel that were actually found and measured on both
  frames.  Finally, the \emph{right panel} shows the \emph{measured} photometry
  for the stars that were recovered.  The heavy dashed lines in the middle and
  right panels show the 50\% detection completeness levels.  }
\label{cmd_fake}
\end{figure*}

To test the internal errors and completeness of the photometry we ran two
separate artificial-star procedures.  In the first series, stars were added to
a representative $1500 \times 500-$px region of the image in the lower left
corner (similar to the region shown in Figure 1).  These added stars were
distributed evenly in color and magnitude over the ranges $0.5 < (V-I) < 5.0$
and $27.0 < I < 29.6$, as shown in Figure \ref{cmd_fake}.  These intervals
deliberately covered a larger range in both color and magnitude than in our
observed CMD (Figure \ref{cmd4}).  The same \emph{daophot} measurement sequence
as done on the original frames was then carried out, with a two-pass sequence
of \emph{find/phot/allstar}.

\begin{figure*}
\centering
\includegraphics[height=7.0cm]{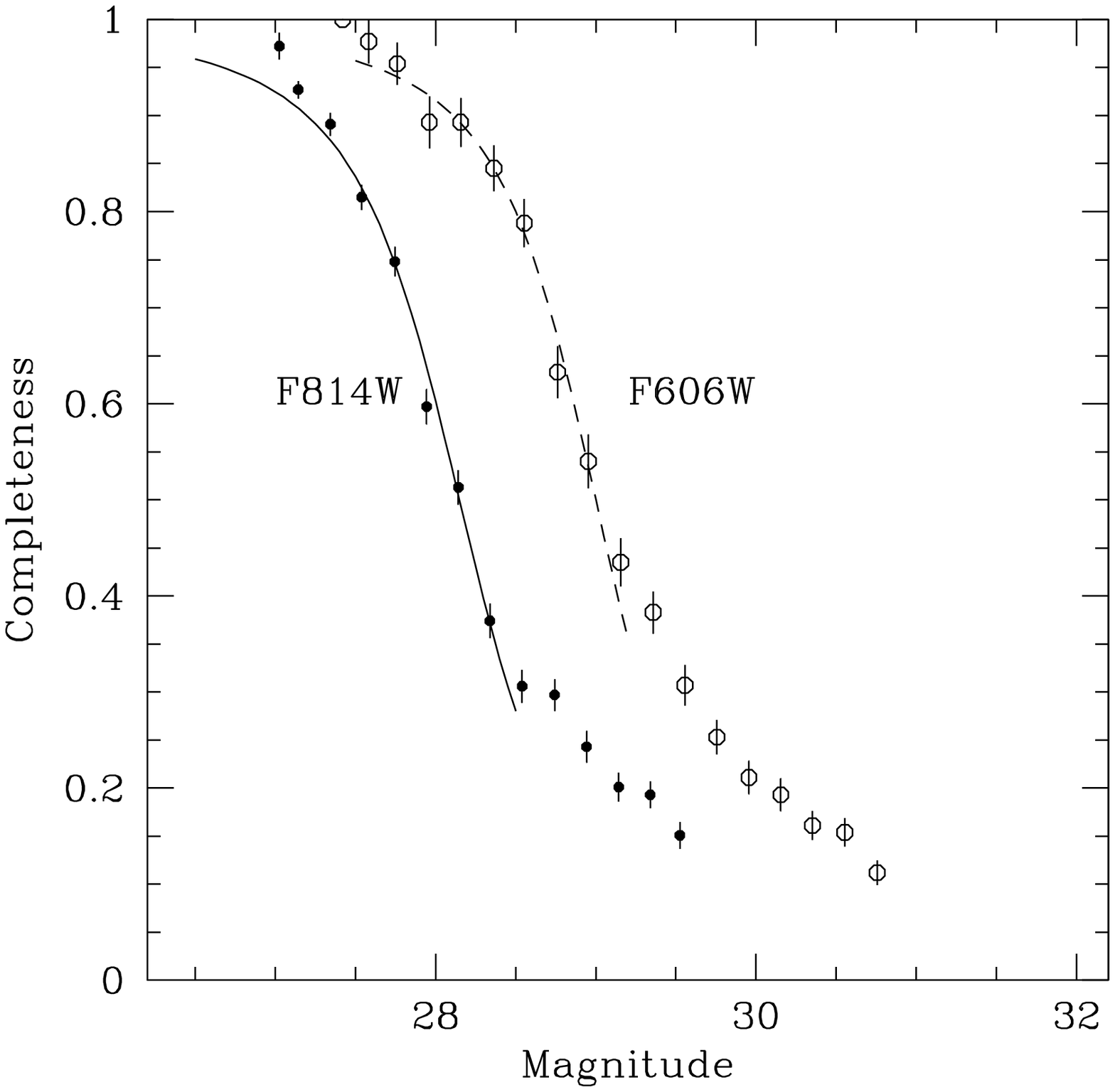}
\caption{The photometric completeness as determined from the artificial-star
  tests in the previous figure.  The ratio $f$ represents the fraction of input
  stars actually found and measured in both colors, and is plotted here as a
  function of magnitude in each filter.  The curves drawn through the brighter
  parts of each set of points is a standard interpolation function of the form
  $f = \frac{1}{2}[1-{\alpha(m-m_0)}/\sqrt{1+\alpha^{2}(m-m_0)^{2}}]$ where
  $m_0$ is the 50\% completeness level and $\alpha$ governs the steepness of
  the falloff.  }
\label{completeness}
\end{figure*}

Figure \ref{cmd_fake} shows the results for inserted stars that were actually
recovered in the photometry, including both their input magnitudes and colors
(center panel) and their actually measured values (right panel).  Of the total
of 10000 added stars, just 1917 of these were successfully found and measured
in both colors, and a high fraction of these lie in the upper left (bright,
blue) part of the CMD.  The completeness of detection $f =
n(recovered)/n(input)$ as a function of magnitude is shown in Figure
\ref{completeness}; the 50\% completeness levels are reached at $F606W = 29.00$
and $F814W = 28.15$.  To be classified as ``recovered'' a star must be detected
in both images.  Note that at very faint levels (below the 50\% point) the
formal values of $f$ tend to decrease rather slowly, a result of the very high
degree of crowding.

\begin{figure*}
\centering
\includegraphics[height=8.0cm]{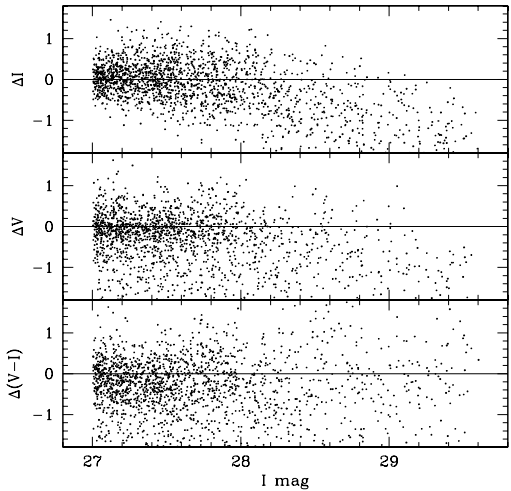}
\caption{Random uncertainties in the photometry, determined from the first
  series of artificial-star tests described in the text.  The first and second
  panels show the differences $\Delta I$, $\Delta V$ (measured \emph{minus}
  input) for the recovered stars, while the third panel shows the difference in
  color.  }
\label{random}
\end{figure*}

Figure \ref{cmd_fake} (particularly the difference between the second and third
panels) clearly indicates that internal random uncertainties are large at all
levels of the CMD.  Figure \ref{random} displays the differences between the
measured and input magnitudes more completely.  For $I \gtrsim 28.0$, one
magnitude below the RGB tip, completeness of detection becomes low and the
systematic errors increase, such that most stars are measured too bright.  We
do not consider this faint region further.  The measured \emph{colors} are on
average slightly too blue independent of magnitude, by $\langle \Delta (V-I)
\rangle = -0.22$ mag.  Over the top magnitude of the RGB ($I = 27 - 28$) the
internal random scatter of the magnitudes and colors (rms) is $\sigma =
\pm0.36$ mag in $F814W$, $\pm0.53$ mag in $F606W$, and $\pm 0.40$ mag in
$(F606W-F814W)$.

The artificially even distribution of input stars in Figure \ref{cmd_fake}
overpopulates the brightest part of the luminosity function compared with the
real CMD.  A second run of artificial-star testing was therefore done, this
time by inserting a set of RGB stars that followed a real CMD for the halo of a
giant elliptical.  We took the photometry for the halo stars of NGC 5128
measured by Rejkuba et al. (\cite{rej05}) with deep HST/ACS exposures that
reach the horizontal branch and thus accurately populate the entire luminosity
and color range of interest here.  Of the more than 77000 stars in the Rejkuba
et al. dataset, about half (38640) were chosen randomly; these were shifted to
the distance and reddening of M87 to ensure that they occupied the same range
in intrinsic luminosity and color; and were assigned random (x,y) locations in
the same $1500\times500$ section of the M87 image as before.  These added stars
cloned from NGC 5128 increased the total light in the image section by less
than 2\%.

\begin{figure*}
\centering
\includegraphics[height=8.0cm]{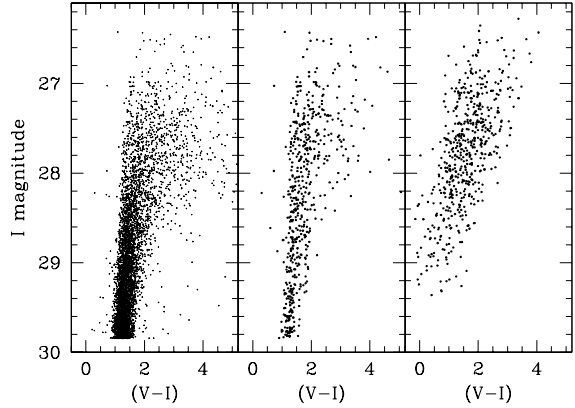}
\caption{Results of the second artificial-star set of tests.  \emph{Left panel}
  shows the color-magnitude distribution of the artificial stars inserted into
  the image section as described in Section 5.  The stars are taken from the
  NGC 5128 halo field measured by Rejkuba et al. (\cite{rej05}), shifted to the
  distance and reddening of M87.  \emph{Middle panel} shows the stars from the
  left panel that were actually found and measured on both frames; and the
  \emph{right panel} shows the \emph{measured} photometry for the stars that
  were recovered.  }
\label{cmd5128}
\end{figure*}

The results of the remeasurement process are shown in Figure \ref{cmd5128}.  Of
5200 input stars brighter than $I=29.8$, just 570 were recovered after the
2-pass photometry.  Interestingly, the faintest ones recovered ($I \gtrsim 29$)
often have very large systematic errors that push them well upward in the CMD.
However, the region $I < 28$ preserves much useful structure representing the
metallicity range, losing only the reddest part of the range.

\section{The Metallicity Distribution}

The tests of the photometry described above lead us to restrict the MDF
analysis to the brightest part of the RGB at $28.0 > I > 27.0$.  In this range,
even the random $\pm 0.40-$mag scatter in $(V-I)$ is much smaller than the
1.5-mag intrinsic spread in the actual CMD, enough to give us workable leverage
on the underlying metallicity distribution.  In addition, the $\simeq 0.2-$mag
systematic error in the colors is far smaller than the intrinsic color range of
the red-giant tracks as they fan out toward the top of the RGB (Figure
\ref{cmd_fiducials}), creating a shift at worst of 0.1-0.2 dex in mean
metallicity.

To estimate the combined effects of the quality of photometry on the deduced
MDF, we used the NGC 5128 data (Figure \ref{cmd5128}) again as a testbed.  With
the procedures described in detail in previous papers (Harris et al.
\cite{har02}, \cite{har07}, Rejkuba et al. \cite{rej05}), interpolation within
the grid of RGB tracks yields the MDF in the form of number of stars per unit
interval in [m/H] = log$(Z/Z_{\odot})$.  The inner halo stars within 8 kpc in
NGC 5128 are now well established to have a broad, predominantly metal-rich MDF
peaking near [m/H] $\simeq -0.3$, and other giant ellipticals are expected to
be similar (see the extensive discussions in Harris et al. \cite{har02},
Rejkuba et al. \cite{rej05} among others).

\begin{figure*}
\centering
\includegraphics[height=8.0cm]{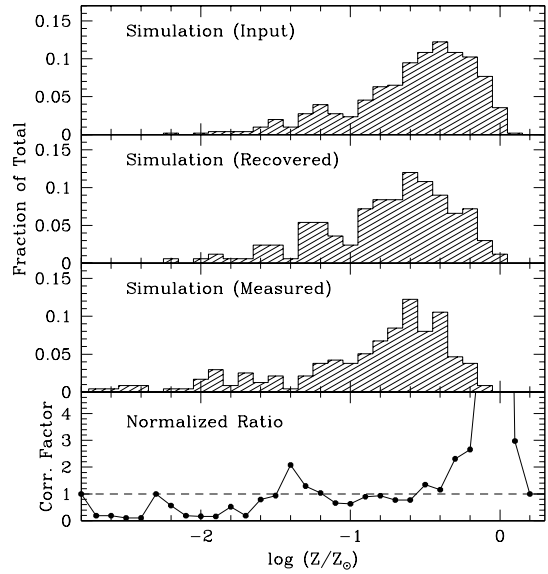}
\caption{Calibration tests of the metallicity distribution.  In the top panel,
  the MDF is shown for the all the NGC 5128 stars from $I = 27$ to 28 from the
  first panel of the previous figure.  In the second panel, the MDF is shown
  for all the NGC 5128 stars that were \emph{recovered} in the simulation tests
  (those in the middle panel of the previous figure).  In the third panel, the
  MDF for these same recovered stars is shown, but now derived from their
  photometry as actually measured from the simulations (i.e. from the third
  panel in the previous figure). All three panels are normalized to the same
  number of stars. The bottom panel shows the ratio of the first and third
  histograms, normalized to an overall average of 1.00.  }
\label{simhisto}
\end{figure*}

The RGB-grid interpolation procedure applied to the NGC 5128 stars is shown in
three different versions in Figure \ref{simhisto}.  In the top panel, we take
the 507 stars in Figure \ref{cmd5128}(a) that lie in our adopted range $27 < I
< 28$ and derive the MDF in histogram form as shown there.  This can be thought
of as the intrinsic, input distribution free of photometric selection or
scatter.

In the second panel, the MDF derived through the interpolation grid is shown
for only the 167 NGC 5128 stars that were \emph{recovered} in the simulation
tests (that is, those in Figure \ref{cmd5128}(b)).  The $(I, V-I)$ values
entered into the RGB grid are their true (input) values.  The visible change in
the MDF compared with the intrinsic version is the direct effect of the
$V-$band cutoff, which is responsible for the loss of the highest-metallicity
stars and thus also decreases the metallicity of the histogram peak.

In the third panel, the MDF is now shown for the recovered stars at $R_{gc} =
115''-155''$ or $9.3 - 12.5$ kpc between $I=27 - 28$ derived from their
actually measured photometry (Figure \ref{cmd5128}(c)).  The overall histogram
range and peak remain similar to the second panel, but relatively more stars
appear at low metallicity because of the systematic bias of the colors toward
the blue as mentioned above.

Lastly, the bottom panel of Figure \ref{simhisto} shows the ratio of the first
and third histograms (input versus measured), normalized to an overall average
of 1.00.  We use this ratio to make a reasonable correction of the MDF for both
the incompleteness (which affects mainly the high-metallicity end) and the
color bias (which affects mainly the low-metallicity end).

\begin{figure*}
\centering
\includegraphics[height=9.0cm]{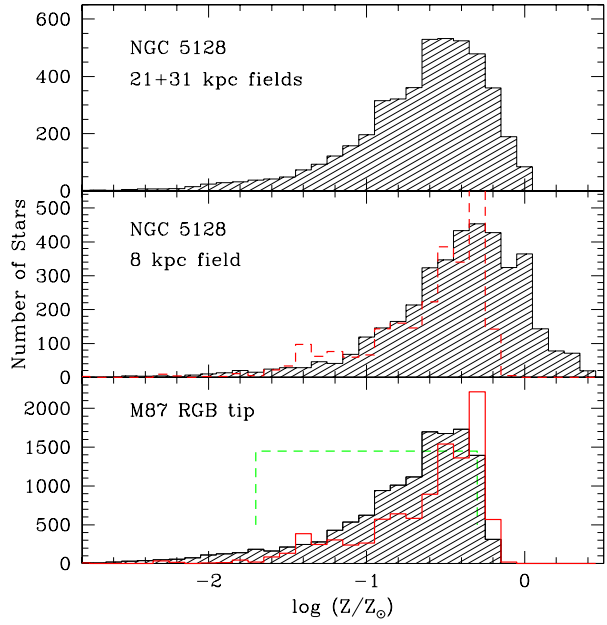}
\caption{Metallicity distribution for the halo stars in NGC 5128 and M87.  The
  upper two panels show the MDF (number of stars per 0.1-dex metallicity
  interval) from three locations in the halo of NGC 5128 (see text).  In the
  bottom panel, the shaded histogram shows the MDF for the stars in the M87
  inner halo in the magnitude range $I = 27.0-28.0$, uncorrected for
  photometric completeness or color bias.  The heavy red \emph{solid line}
  shows the MDF now fully corrected for those effects, while the heavy green
  \emph{dashed line} shows the metallicity range in which the deduced MDF is
  most reliable.  This corrected MDF for M87 is repeated in the middle panel
  (dashed line) for direct comparison with the inner-halo field in NGC 5128
  (vertical scaling is arbitrary for display purposes).  }
\label{fehhisto}
\end{figure*}

The final results for our estimated MDF are shown in Figure \ref{fehhisto}.
The 13972 M87 stars from Figure \ref{cmd_fiducials} in the range $I=27.0-28.0$
as described above are put through the interpolation grid of RGB tracks to
construct the raw MDF shown in the bottom panel (shaded histogram).  This
distribution is then multiplied by the correction factor shown in the bottom
panel of Figure \ref{simhisto} to produce the \emph{fully corrected MDF} shown
as the heavy unshaded line.  This corrected distribution represents our best
estimate of the underlying MDF for the inner halo of M87, corrected for
photometric incompleteness and color measurement bias.

Even this corrected MDF can be relied on only for [m/H] $\lesssim -0.3$; any
stars more metal-rich than that are almost totally cut off from our data and
thus no correction factors are useful.  At the opposite end, the MDF shape for
[m/H] $\lesssim -1.7$ is very poorly determined and we can say only that there
are very few stars present in that range.

For direct comparison purposes we also show MDF data for the halo stars in
three locations of the NGC 5128 halo, from Harris et al.  (\cite{har99},
\cite{har00}, \cite{har02}).  These were all taken with the HST WFPC2 camera in
$(V,I)$ and have rather similar limiting absolute magnitudes to our M87
ACS-based photometry, though they are less affected by crowding.  The MDFs were
derived in all cases with the same RGB grid of tracks and interpolation code.
The mid-to-outer fields at projected galactocentric distances of 21 and 31 kpc
have MDFs that are virtually identical to each other and are combined in the
upper panel of Figure \ref{fehhisto}, while the inner-halo 8 kpc field is shown
by itself in the middle panel.

Our corrected MDF for the M87 inner halo -- at a mean projected distance of 10
kpc -- clearly resembles the 8-kpc NGC 5128 field more closely than the outer
fields.  Over the range [m/H] $< -0.3$ where we can make the comparison, we
conclude that the inner halos of both these giant ellipticals have basically
similar MDFs that are broad, predominantly metal-rich, and with very small
numbers of metal-poor stars.  The MDF for M87 \emph{may} reach a peak near
[m/H] $\sim -0.4$, but for the present this metallicity should be viewed as a
lower limit to the peak value.


\section{Conclusions}

We have used HST Archive data for an unusually deep set of $F606W, F814W$ ACS
images to probe the red-giant stellar population in the inner halo of M87.
Although the crowding levels of the faint halo stars in this field are severe
at best, the regions from $115'' - 155''$ (9.3 to 12.5 kpc projected
galactocentric distance) give a useful first look at the brightest 1.5
magnitudes of the red-giant branch. This corresponds to distances of $1.4 R_e -
1.9 R_e$, where $R_e$ is the effective radius in the I band of 81'' (6.3 kpc)
(Zeilinger, \cite{zei93}).

We have used this material to obtain a preliminary TRGB-calibrated distance to
M87 for the first time.  We find a distance $d = (16.7 \pm 0.9)$ Mpc, in good
agreement with the few other relatively direct methods available, including the
planetary nebula luminosity function, the Cepheid-calibrated SBF method, and
the linear diameters of globular clusters.  Photometry of a more outlying halo
field than the one we are able to study here has the potential to give a more
precise TRGB determination and would represent the most direct possible
distance calibration to the Virgo members at the present time.

As a second result of our study, we use the brightest 1 magnitude of the RGB
stars to measure the metallicity distribution of the M87 inner halo for the
first time.  We find that it is very broad and predominantly metal-rich, with a
peak that may lie very near the completeness limit of our $V-$band data but is
at least as metal-rich as [m/H] $\simeq -0.4$.  In general, the shape of the
MDF strongly resembles the ``8-kpc field'' for the inner halo of NGC 5128, the
nearby giant elliptical in the Centaurus group.  In our view, these results
strongly reinforce the view that the oldest stellar populations in giant E
galaxies are fundamentally similar, regardless of whatever more recent
accretion events may have happened to them.

Our study and the Williams et al. (\cite{wil07}) photometry of an intracluster
Virgo field, show that there is considerable potential for extending this type
of work to other Virgo members and the intracluster population as well.  HST
imaging with the ACS and WFC3 cameras is easily capable of resolving their halo
red-giant populations and tracing the metallicity structures in this rich,
nearby collection of large galaxies.

\begin{acknowledgements}
SB acknowledges the Wihurin Foundation and the V\"{a}si\"{a}l\"{a} Foundation
for financial support. WEH acknowledges support from the Natural Sciences and
Engineering Research Council of Canada.
\end{acknowledgements}

\end{document}